\documentclass[aps,prl,preprint,showpacs,showkeys,a4paper]{revtex4-1}

\usepackage{graphicx}
\usepackage{dcolumn}
\usepackage{amsmath,amssymb,mathtools}
\usepackage{epstopdf}
\usepackage{verbatim}

\DeclareMathOperator\arctanh{arctanh}
\newtheorem{theorem}{Theorem}

\begin{document}


\title{Universal cosmological solutions in Lovelock gravity}



\author{A. V. Nikolaev}\email[]{ilc@xhns.org}
\affiliation{Ulyanovsk State Pedagogical University,  Ploschad Lenina 4/5, Ulyanovsk, Russia}

\date{\today}

\begin{abstract}
This paper explores the Friedmann field equations within the framework of Lovelock gravity, a natural extension of Einstein's gravity, focusing on both flat and open universes. Utilizing an approach based on independent Riemann tensor components, we derive generalized Friedmann equations for Lovelock gravity and categorize the solutions into Type I and Type II types. We identify additional vacuum solutions in a flat universe and present a comprehensive solution for a pressure-free scenario in an open universe, both unique to Lovelock gravity. These findings provide new insights into the cosmological implications of Lovelock gravity and offer a foundation for further exploration into the universe's evolutionary trajectory.
\end{abstract}


\maketitle


\section{Introduction}
A. Friedmann authored two of his most pivotal papers in the realm of general relativity in 1922 and 1924. These works, published in the German journal Zeitschrift für Physik, marked the advent of non-stationary solutions to Einstein's equations. Initially, Einstein contested the validity of Friedmann's findings, however, he later endorsed them \cite{Fok1963}. This revelation underscored the groundbreaking nature of the concept of an expanding universe during that era. Subsequent observations by E. Hubble further substantiated the non-stationary nature of the universe \cite{Hubble1929}. Hence, Friedmann equations are used extensively in cosmology. Friedmann equations in general relativity are well studied; see for reference \cite{Weinberg2008,Misner2017}. Methods for obtaining exact solutions in Friedmann universes are also well studied; see for reference \cite{StephaniES2003}. Solutions of Friedmann field equations have wide usage in cosmology as background solutions \cite{Plank2016}, in astrophysics for distance measurements \cite{Nikolaev2017}, and in other fields. Thus, such solutions have significant interest in analyzing modified theories of gravity, which have attracted a lot of interest in recent years. In this paper, we will focus on Lovelock gravity.

The study of Lovelock gravity and Einstein-Gauss-Bonnet (EGB) gravity is a hot topic for the world scientific community of theoretical physicists working in the field of gravity theory and cosmology. The reason for this is the fact that Lovelock's gravity is a natural generalization of Einstein's gravity and is one of the main contenders for replacing general relativity (GR). Another reason for the active research of this theory of gravity is its novelty and the high probability of obtaining new, interesting results.

In EGB gravity, several exact solutions have been found for specific stellar models, for example \cite{Davis2003,Dadhich2010,Kang2012,Chilambwe2015,Hansraj2015,Maharaj2015,Hansraj2019,Bhar2017,Hansraj2017,Sardar2019}. In paper \cite{Wright2016}, the Buchdahl limit was found for the isotropic distribution of matter, which made it possible to qualitatively analyze black holes in the EGB theory. Several interesting results in the study of cosmological problems in the EGB theory have been reported \cite{Andrew2007,Guo2009,Mukerji2010,Canfora2018,Armaleo2018,Diaz2020a}.

In most cases, research in the field of EGB gravity is of a particular nature. Overall results are extremely rare. Among such results, one can single out the work of Pavluchenko and Toporensky \cite{Pavluchenko2018}, in which the effects of spatial curvature and anisotropy in asymptotic regions are studied. In the paper by S. Capozziello et al \cite{Capozziello2019}, it is shown that the Gauss-Bonnet invariant can be associated with an ideal fluid. Diaz et al. \cite{Diaz2020b} showed that a brane-type metric is possible for models of an accelerating universe. N. Dadhich et al \cite{Dadhich2015} obtained solutions describing black holes.

In the case of Lovelock gravity, the results are fewer, since this theory is more general and more complex than the EGB gravity, but the results obtained in Lovelock gravity are more general and applicable to the EGB gravity. The vacuum models were obtained in the works \cite{Wheeler1986a,Wheeler1986b,Wiltshire1986,Myers1988,Banados1994,Banados1992}.

Solutions for black holes in spaces affected by string theory were found by N. Dadhich et al \cite{Dadhich2013}. Toledo and Bezerra \cite{Toledo2019} found a solution for a spherically symmetric black hole with quintessence in pure Lovelock gravity. The generalized Vaidya space has been used to study gravitational collapse and generalized Misner-Sharp energy in the following works \cite{Cai2008,Rudra2011,Dadhich2013B,Nozawa2006}.

Choosing a specific gravitational potential can greatly simplify the finding of solutions in Lovelock gravity. Thus, solutions were obtained for the global monopole and the BTZ black hole \cite{Dadhich2012}, for the Nariai and Bertotti spaces \cite{Dadhich2013C,Batista2016}, Gödel Universe \cite{Dadhich2017}, and Kasner metrics \cite{Camanho2015}.

The distribution of matter in the case of static spherically symmetric spaces was considered for the case of isothermal spheres \cite{Dadhich2016}, for the Vaidya-Tikekar geometry \cite{Khugaev2016}. Static spheres in the form of compact objects were considered in the work of N. Dadhich et al \cite{Dadhich2017B}. The limits of compactness of relativistic stars with electromagnetic effects are considered in the work of S. Chakraborty et al \cite{Chakraborty2020}.

It will be interesting to note the work of M. Gurses and Y. Heydarzade \cite{Gurses2020}, in which an approach to solving cosmological problems in generalized theories of gravity is considered in general form. The results obtained in this work are close to the tasks set in the proposed project, but they are of a particular nature, since only cosmological metrics are considered in it.

In this paper, we are using the independent Riemann tensor components approach to obtain the general form of Friedmann field equations. The main goal of this paper is to provide a toolkit to study Friedmann field equations in Lovelock gravity in general. In section \ref{sec1}, we explain the used approach, in section \ref{sec2}, we obtained the general form of Friedmann equations in Lovelock gravity. In section \ref{sec3}, we discuss possible Type I and Type II solutions for the obtained equations.

\section{The approach}
\label{sec1}
The approach being proposed extends Karmarkar's method, through which he derived the embedding condition for Karmarkar class I~\cite{Karmakar1948}. In his paper, he delineated the independent components of the Riemann metric for spherically symmetric spacetime and, by applying the Gauss-Codazzi-Ricci equations, arrived at the well-known Karmarkar's condition.
\begin{equation}
	R_{1220}R_{1330}+R_{1010}R_{2323} - R_{1212}R_{3030} = 0.
	\label{karmarkar}
\end{equation}

Later, Prasanna demonstrated that this condition remains invariant under arbitrary non-singular Gaussian transformations that preserve spherical symmetry, and it can be reformulated as shown below:
\begin{equation}
    \frac{f_1f_4 - (f_5)^2}{f_2f_3} = I.
    \label{prasana}
\end{equation}
Here, $f_i$ represent the independent components of the Riemann tensor. This result was presented in the work~\cite{Prasanna1969}. Tikekar~\cite{Tikekar1970} utilized the method of independent Riemann tensor components to analyze a spherically symmetric metric belonging to class one. The analysis underscored the specific conditions that must be met concerning the Riemann tensor components for a perfect fluid distribution that is not conformally flat.

Below, we will demonstrate how to extend this approach to $N$-dimensional spherically symmetric spacetime and apply it to Lovelock gravity.

We utilize $N$-dimensional spherical symmetric coordinates $[t, r,\phi_1,\phi_2,\dots,\phi_{N-2}]$. Latin indices represent spacetime coordinates $a=0..N$, while Greek indices represent angle coordinates $\alpha = 2..N-2$. For convenience, we set $c = 8\pi G = 1$ in this paper. The Friedmann-Robertson-Walker-Lemaître (FRWL) metric we will be working with is given by
\begin{equation}
    ds^2 = -dt^2 + a(t)^2\left(\frac{1}{1 - kr^2}dr^2 + r^2d\Omega_{N-2}^2\right),
    \label{metricN}
\end{equation}
where $d\Omega_{N-2}^2$ represents the angular coordinates in the $N-2$ dimensional sphere.

Let's analyze the structure of the Riemann tensor for the metric in Eq. \eqref{metricN}. We will now express all its non-zero components for $N=4$, while omitting components that arise from the symmetries of the Riemann tensor.

\begin{subequations}
    \begin{equation}
        (N = 4):\ \   R^{12}_{\ \ 12} = R^{13}_{\ \ 13} = R^{23}_{\ \ 23} = \frac{\dot{a}^2}{a^2} + \frac{k}{a^2},
    \end{equation}
    \begin{equation}
        (N = 4):\ \ R^{10}_{\ \ 10} = R^{20}_{\ \ 20} = R^{30}_{\ \ 30} = \frac{\ddot{a}}{a},
    \end{equation}
\end{subequations}
now we will do the same for $N=5$
\begin{subequations}
    \begin{equation}
        (N = 5):\ \   R^{12}_{\ \ 12} = R^{13}_{\ \ 13} = R^{14}_{\ \ 14} = R^{23}_{\ \ 23} = R^{24}_{\ \ 24} = R^{34}_{\ \ 34} = \frac{\dot{a}^2}{a^2} + \frac{k}{a^2},
    \end{equation}
    \begin{equation}
        (N = 5):\ \ R^{10}_{\ \ 10} = R^{20}_{\ \ 20} = R^{30}_{\ \ 30} = R^{40}_{\ \ 40}  = \frac{\ddot{a}}{a}.
    \end{equation}
\end{subequations}
Thus, we observe that the Riemann tensor has only 2 independent components, which we can express generally as:
\begin{subequations}
    \begin{align}
        \label{notation5_f}
        f_1 =& R^{1\alpha}_{\ \ 1\alpha} = R^{\alpha\beta}_{\ \ \alpha\beta} = \frac{\dot{a}^2}{a^2} + \frac{k}{a^2},\\
        f_2 =& R^{10}_{\ \ 10} = R^{\alpha 0}_{\ \ \alpha 0} = \frac{\ddot{a}}{a}.
        \label{notation5_l}
    \end{align}
\end{subequations}

It should be noted that throughout this discussion, unless stated explicitly, $\alpha < \beta$. All other components of the Riemann tensor can be derived from equations \eqref{notation5_f} and \eqref{notation5_l}. The detailed mathematical proof of this assertion can be found in Appendix \ref{app1}.

This notation, which is based on the independent components of the Riemann tensor, allows us to analyze the Lovelock field equations in a simpler and more transparent manner.

\subsection{Lovelock Gravity}

Lovelock gravity (LG) serves as the natural extension of Einstein gravity. Therefore, the Friedmann equations in General Relativity (GR) should find their counterpart in Lovelock gravity. In this discussion, we adopt slightly modified notations from Dadhich \cite{Dadhich2010B}, utilizing the more common generalized Kronecker delta notation instead of the determinant symbol. 

The field equations in LG are represented as follows:
\begin{equation}
    \mathcal{\prescript{(1)}{}H}_{ab} + \sum_{i=2}^{n}\alpha_i \mathcal{\prescript{(i)}{}H}_{ab} = T_{ab},
    \label{Lavelockfield}
\end{equation}
where $\alpha_i$ denotes coefficients, $\prescript{(n)}{}H_{ab}$ signifies the Lovelock $n$-order polynomial defined by:
\begin{equation}
    \mathcal{\prescript{(n)}{}H}_{ab} = n \mathcal{R}_{ab} - \frac{1}{2}\mathcal{R}g_{ab},
    \label{LHDef}
\end{equation}
Here, $\mathcal{R}_{ab}$ and $\mathcal{R}$ represent the generalized Ricci tensor and Ricci scalar respectively:
\begin{subequations}
    \begin{equation}
        \mathcal{R}_{ab} = g^{cd}\mathcal{R}_{cadb},
        \label{LRicciDef}
    \end{equation}
    \begin{equation}
        \mathcal{R} = g^{ab}\mathcal{R}_{ab}, 
        \label{LRicciSDef}
    \end{equation}
\end{subequations}      
where $\mathcal{R}_{abcd}$ denotes the generalized Riemann tensor:
\begin{equation}
    \mathcal{R}_{abcd} = Q_{ab}^{\ \ mn}R_{mncd},
    \label{PureRdef}
\end{equation}
where $Q^{ab}_{\ \ cd}$ is defined as:
\begin{equation}
    Q^{ab}_{\ \ cd} = \frac{1}{2^n}\delta^{aba_2b_2\dots a_{n} b_{n}}_{cdc_2d_2\dots c_{n} d_{n}} R_{a_2b_2}^{\ \ c_2d_2}\dots R_{a_{n}b_{n}}^{\ \ c_{n}d_{n}},
    \label{PureQdef}
\end{equation}
and $\delta^{aba_2b_2\dots a_{n} b_{n}}_{cdc_2d_2\dots c_{n} d_{n}}$ represents the generalized Kronecker delta.

\section{Friedmann equations in Lovelock gravity}
\label{sec2}
Based on the information provided in section \ref{sec1}, it is understood that in the case of the FRWL metric, the Riemann tensor is characterized by only two independent components, denoted as $f_1$ and $f_2$. Leveraging this knowledge, we can express the field equations of Lovelock gravity in a more simplified manner. Reference to the literature (such as \cite{Dadhich2010B}) indicates that only Lovelock gravities where $N \geq 2n+1$ hold physical significance. By considering this criterion, we can avoid the need for employing the Heaviside step function in calculations, which typically arises in more general cases.

Let's begin with the analysis of $Q^{ab}_{\ \ cd}$, which also exhibits only two independent components
\begin{subequations}
    \begin{align}
        \label{fQ1}
	Q^{1\alpha}_{\ \ 1\alpha} = Q^{\alpha\beta}_{\ \ \alpha\beta} &= \frac{\left(2\left( n - 1 \right)\right)!}{2}\left[ C_{N - 3}^{2\left( n - 1 \right)}f_1 + C_{N - 3}^{2\left( n - 1 \right) - 1}f_2 \right]f_1^{n - 2} = q_1,\\
	Q^{10}_{\ \ 10} = Q^{\alpha 0}_{\ \ \alpha 0} &= \frac{\left(2\left( n - 1 \right)\right)!}{2}C^{2\left( n - 1 \right)}_{N - 2}f_1^{n - 1} = q_2,
        \label{fQ2}
    \end{align}
\end{subequations}
where $C^k_n =\frac{n!}{\left( n - k \right)!k!}$. Please note that the subsequent general expressions for $Q^{ab}_{\ \ cd}$ are presented for the case where $n > 1$. This choice is made to ensure simplicity and brevity in the expressions.

The generalized Riemann tensor components can be derived from \eqref{PureRdef}, yielding the following independent components:
\begin{subequations}
    \begin{align}
        \mathcal{R}^{1\alpha}_{\ \ 1\alpha} = \mathcal{R}^{\alpha\beta}_{\ \ \alpha\beta} &= 2Q^{12}_{\ \ 12}R^{12}_{\ \ 12} = 2 q_1f_1,\\
        \mathcal{R}^{10}_{\ \ 10} = \mathcal{R}^{\alpha 0}_{\ \ \alpha 0} &= 2Q^{10}_{\ \ 10}R^{10}_{\ \ 10} = 2 q_2f_2.
    \end{align}
\end{subequations}

Next, by utilizing \eqref{LRicciDef} and \eqref{LRicciSDef}, we can determine the components of the generalized Ricci tensor and the generalized Ricci scalar:
\begin{subequations}
    \begin{align}
        \mathcal{R}^{1}_{1} = \mathcal{R}^\alpha_\alpha &= \mathcal{R}^{10}_{\ \ 10} + \left( N - 2 \right)\mathcal{R}^{12}_{\ \ 12} = 2\left(q_2f_2 + \left( N - 2 \right)q_1f_1\right),\\
        \mathcal{R}^{0}_{0} &= \left( N - 1 \right)\mathcal{R}^{10}_{\ \ 10} = 2\left( N -1 \right)q_2f_2,\\
        \mathcal{R} &= \mathcal{R}^{0}_0 + \left( N - 1 \right)\mathcal{R}^1_1 = 4\left( N - 1 \right)q_2f_2 + 2\left( N - 1 \right)\left( N - 2 \right)q_1f_1.
    \end{align}
\end{subequations}
Finally, from \eqref{LHDef}, the Lovelock tensor components are given by:
\begin{subequations}
    \begin{align}
        \mathcal{\prescript{(n)}{}H}^{1}_{1} = \mathcal{\prescript{(n)}{}H}^{\alpha}_{\alpha} &= n\mathcal{R}^{1}_{1} - \frac{1}{2}\mathcal{R} = -\left( N - 2 \right)\left( N - 1 - 2n \right)q_1f_1 - 2\left( N - n - 1 \right)q_2f_2,\\
        \mathcal{\prescript{(n)}{}H}^{0}_{0} &= n\mathcal{R}^{0}_{0} - \frac{1}{2}\mathcal{R} = -\left( N - 1 \right)\left[ \left( N - 2 \right)q_1f_1 - 2\left( n - 1 \right)q_2f_2 \right].
    \end{align}
\end{subequations}

By utilizing the expression for $Q^{ab}_{\ \ cd}$ specified in \eqref{fQ1} and \eqref{fQ2}, we can derive the following expressions:
\begin{subequations}
    \begin{equation}
	    \mathcal{\prescript{(n)}{}H}^{1}_{1} =  f_1^{n-1}\left(\prescript{(n)}{}k_{11}f_1 + \prescript{(n)}{}k_{12} f_2\right),
        \label{LHf1}
    \end{equation}
    \begin{equation}
        \mathcal{\prescript{(n)}{}H}^{0}_{0} =  f_1^{n-1}\left(\prescript{(n)}{}k_{21}f_1 + \prescript{(n)}{}k_{22} f_2\right),
        \label{LHf2}
    \end{equation}
\end{subequations}

Where the coefficients are defined as follows:
\begin{subequations}
    \begin{align}
        \label{eqK11}
	    \prescript{(n)}{}k_{11} &= - \frac{1}{2}\left(2\left( n - 1 \right)\right)!\left( N - 2 \right)\left( N - 1 -2n \right)C^{2\left( n - 1 \right)}_{N - 3},\\
	    \prescript{(n)}{}k_{12} &= -\frac{1}{2}\left(2\left( n - 1 \right)\right)!\left[ 2\left( N - n - 1 \right)C^{2\left( n - 1 \right)}_{N - 2} + \left( N - 2 \right)\left( N - 1 -2n \right)C^{2n - 3}_{N - 3} \right],\\
	    \prescript{(n)}{}k_{21} &= - \frac{1}{2}\left(2\left( n - 1 \right)\right)!\left( N - 1 \right)\left( N - 2 \right)C^{2\left( n - 1 \right)}_{N - 3},\\
	    \prescript{(n)}{}k_{22} &= \frac{1}{2}\left(2\left( n - 1 \right)\right)!\left( N - 1 \right)\left[ 2\left( n - 1 \right)C^{2\left( n -1 \right)}_{N - 2} - \left( N - 2 \right)C^{2n - 3}_{N - 3} \right] = 0.
        \label{eqK22}
    \end{align}
\end{subequations}

An intriguing observation is that the coefficients $\prescript{(n)}{}k_{11}$, $\prescript{(n)}{}k_{12}$, and $\prescript{(n)}{}k_{21}$ are algebraically related as shown in the equation:
\begin{equation}
    \boxed{
    \prescript{(n)}{}k_{11} + \prescript{(n)}{}k_{12} - \prescript{(n)}{}k_{21} = 0
}
    \label{KMagic}
\end{equation}
This relationship will be demonstrated in subsequent sections and proves to be a very useful identity.

Utilizing \eqref{LHf1} and \eqref{LHf2} along with \eqref{Lavelockfield}, we can express the generalized Friedmann equations in terms of independent Riemann tensor components:
\begin{subequations}
    \begin{equation}
        -\left( N - 2 \right)\left( \frac{N - 3}{2}f_1 + f_2 \right) + \sum_{i=2}^n \alpha_i f_1^{i-1}\left( \prescript{(n)}{}k_{11}f_1 + \prescript{(n)}{}k_{12}f_2 \right) = p,
        \label{RFriedman1}
    \end{equation}
    \begin{equation}
        -\frac{1}{2}\left( N - 1 \right)\left( N - 2 \right)f_1 + \sum_{i=2}^n \alpha_i f_1^{i}\prescript{(n)}{}k_{21} = -\rho.
        \label{RFriedman2}
    \end{equation}
\end{subequations}
Alternatively, expressing these equations in terms of \eqref{metricN}, we have:
\begin{subequations}
    \begin{multline}
        -\left( N - 2 \right)\left( \frac{\left( N - 3 \right)}{2}\left( \frac{\dot{a}^2}{a^2} + \frac{k}{a^2} \right) + \frac{\ddot{a}}{a} \right)+\\
        + \sum_{i=2}^{n}\alpha_i\left[ \prescript{(n)}{}k_{11}\left(  \frac{\dot{a}^2}{a^2} + \frac{k}{a^2} \right)^i + \prescript{(n)}{}k_{12} \frac{\ddot{a}}{a}\left(  \frac{\dot{a}^2}{a^2} + \frac{k}{a^2} \right)^{i-1} \right] =  p,
        \label{gFriedman1}
    \end{multline}
    \begin{equation}
        -\frac{1}{2}\left( N - 1 \right)\left( N - 2 \right)\left( \frac{\dot{a}^2}{a^2} + \frac{k}{a^2} \right)+ \sum_{i=2}^{n}\alpha_i \prescript{(n)}{}k_{21} \left(  \frac{\dot{a}^2}{a^2} + \frac{k}{a^2} \right)^{i}  =  -\rho.
        \label{gFriedman2}
    \end{equation}
\end{subequations}
In specific applications, it is beneficial to employ the equation of motion derived from \eqref{gFriedman1}-\eqref{gFriedman2}:
\begin{multline}
    \left( N-1 \right)\left( N - 2 \right)\frac{\ddot{a}}{a} +\\
    +\sum_{i=2}^n\alpha_i\left[ \left( \prescript{(n)}{}k_{21}\left( N - 3 \right) - \prescript{(n)}{}k_{11}\left( N - 1 \right) \right)\left( \frac{\dot{a}^2}{a^2} + \frac{k}{a^2} \right)^i - \prescript{(n)}{}k_{12}\left( N - 1 \right)\frac{\ddot{a}}{a}\left( \frac{\dot{a}^2}{a^2} + \frac{k}{a^2} \right)^{i-1} \right] =\\
    = -\left( \left( N - 3 \right)\rho + \left( N -1 \right)p \right).
    \label{gFriedmanEM}
\end{multline}
Considering \eqref{eqK11}-\eqref{eqK22}, for the special case when $N=2n+1$, we have $k_{11}=0$ and $k_{12}=k_{21}$, simplifying the generalized Friedmann equations to:
\begin{subequations}
    \begin{equation}
        -\left( N - 2 \right)\left( \frac{\left( N - 3 \right)}{2}\left( \frac{\dot{a}^2}{a^2} + \frac{k}{a^2} \right) + \frac{\ddot{a}}{a} \right) + \sum_{i=2}^{\frac{1}{2}(N-1)}\alpha_i\left[  \prescript{(n)}{}k_{12} \frac{\ddot{a}}{a}\left(  \frac{\dot{a}^2}{a^2} + \frac{k}{a^2} \right)^{i-1} \right] =  p,
        \label{gFriedman1S}
    \end{equation}
    \begin{equation}
        -\frac{1}{2}\left( N - 1 \right)\left( N - 2 \right)\left( \frac{\dot{a}^2}{a^2} + \frac{k}{a^2} \right)+ \sum_{i=2}^{\frac{1}{2}(N-1)}\alpha_i \prescript{(n)}{}k_{12} \left(  \frac{\dot{a}^2}{a^2} + \frac{k}{a^2} \right)^{i}  =  -\rho.
        \label{gFriedman2S}
    \end{equation}
\end{subequations}

\section{Universal solutions}
\label{sec3}
In this section, our objective is to investigate the Friedmann field equations within the framework of Lovelock gravity with the aim of finding solutions that are valid for any Lovelock polynomial order $n$ and spacetime dimensions $N$. We will categorize the solutions into two types:
\begin{itemize}
    \item \textbf{Type I} solution: This solution depends on the specific Lovelock polynomial order, spacetime dimensions, and coupling constants.
    \item \textbf{Type II} solution: This solution does \textit{not} rely on the Lovelock polynomial order, spacetime dimensions, or coupling constants.
\end{itemize}

The importance of such solutions is the fact that they are <<built into>> the theory. For example, Minkowski solution is such a solution in classic GR, if there is no matter present, there is no extra space curvature. In this paper, we focus on such solutions which are provided by extra freedom given by generalization GR to Lovelock gravity.

We will analyze flat, open, and closed universes separately in order to derive solutions without specifying the values of $N$ and $n$. By doing so, we aim to obtain a comprehensive understanding of the Friedmann field equations in the context of Lovelock gravity that transcends specific values of the order and dimensions.

\subsection{Flat universe ($k=0$) solutions}
When focusing on the flat universe case ($k=0$), it is advantageous to reexpress the generalized Friedmann equations \eqref{gFriedman1}-\eqref{gFriedman2} by introducing the Hubble constant $H= \frac{\dot{a}}{a}$ in order to simplify the equations. This reformulation helps in reducing the order of the equations:
\begin{subequations}
    \begin{equation}
        -\left( N - 2 \right)\left( \frac{\left( N - 1 \right)}{2}H^2 + \dot{H} \right)+ \sum_{i=2}^{n}\alpha_i\left[ \left(\prescript{(n)}{}k_{11} + \prescript{(n)}{}k_{12}\right)H^{2i} + \prescript{(n)}{}k_{12}  \dot{H}H^{2i-2} \right] =  p,
        \label{gFriedmanH10}
    \end{equation}
    \begin{equation}
        -\frac{1}{2}\left( N - 1 \right)\left( N - 2 \right)H^2 + \sum_{i=2}^{n}\alpha_i \prescript{(n)}{}k_{21} H^{2i}  =  -\rho.
        \label{gFriedmanH20}
    \end{equation}
\end{subequations}
Or using \eqref{KMagic} 
\begin{subequations}
    \begin{equation}
	    -\left( N - 2 \right)\left( \frac{\left( N - 1 \right)}{2}H^2 + \dot{H} \right)+ \sum_{i=2}^{n}\alpha_i\left[\prescript{(n)}{}k_{12}  \dot{H}H^{2i-2} + \prescript{(n)}{}k_{21}H^{2i} \right] =  p,
        \label{gFriedmanH1}
    \end{equation}
    \begin{equation}
        -\frac{1}{2}\left( N - 1 \right)\left( N - 2 \right)H^2 + \sum_{i=2}^{n}\alpha_i \prescript{(n)}{}k_{21} H^{2i}  =  -\rho.
        \label{gFriedmanH2}
    \end{equation}
\end{subequations}

It is straightforward to observe from equations \eqref{gFriedmanH1}-\eqref{gFriedmanH2} that the trivial vacuum solution is:
\begin{equation}
    H = 0.
\end{equation}
This solution satisfies the vacuum field equations. Alternatively, the solution:
\begin{equation}
    \boxed{a = \mbox{const}.}
    \label{Mink}
\end{equation}
can also be derived. It is evident that the solution \eqref{Mink} represents the \textbf{Type II} solution for Lovelock gravity. From a physical point of view, in this case, we see no difference with classical GR, which means if there is no matter present, we have Minkowski spacetime.

Let's now combine equations \eqref{gFriedmanH1} and \eqref{gFriedmanH1} to obtain:
\begin{equation}
	\dot{H}\left( \left( 2- N \right) +  \sum_{i=2}^{n}\alpha_i\prescript{(n)}{}k_{12}H^{2i-2}  \right) = p+\rho,
	\label{gFriedmanHcomb}
\end{equation}
From equation \eqref{gFriedmanHcomb}, we can observe that there is only one key difference between Lovelock gravity and classic General Relativity in the case of a flat universe - an additional term in the brackets. This extra term gives rise to additional vacuum solutions, such as $H=\text{const}$, specifically:
\begin{equation}
    \label{speccase1}
    H = \text{RootOf} \left( \frac{1}{2} \left( N - 2 \right) \left( N - 1 \right)H^2 - \sum_{i=2}^n\alpha_i \prescript{(n)}{}k_{21}H^{2i} = 0 \right) = \text{const}.
\end{equation}
By leveraging this observation, we can formulate it into the following theorem:
\begin{theorem}
    \label{th1}
    The Lovelock gravity of a polynomial higher than one exhibits (anti-)de Sitter solutions in a flat vacuum universe given by
    \begin{equation}
         H = \text{RootOf} \left( \frac{1}{2} \left( N - 2 \right) \left( N - 1 \right)H^2 - \sum_{i=2}^n\alpha_i \prescript{(n)}{}k_{21}H^{2i} = 0 \right) = C,
	 \label{th1eq}
     \end{equation}
     where $C$ is a constant dependent on $N$ (dimensions), $n$ (Lovelock polynomial order), and $\alpha_i$ (coupling constants).
\end{theorem}

From a physical point of view, the result presented in \eqref{th1eq} has the interesting feature. Namely, generalization of GR to Lovelock gravity leads to existing expanding solutions in vacuum. In other words, it shows that in Lovelock gravity, there is no need to have Dark Energy to explain the inflation.

The solution described in \eqref{speccase1} is dependent on $n$, $N$, and $\alpha_i$, thereby making it a \textbf{Type I} solution.
If we consider the linear equation of state $p = \omega \rho$, we can derive from \eqref{gFriedmanHcomb} the following equation:
\begin{equation}
	\dot{H}\left( \left( 2- N \right) +  \sum_{i=2}^{n}\alpha_i\prescript{(n)}{}k_{12}H^{2i-2}  \right) = \left( 1+\omega \right)\rho.
	\label{gFriedmanHcombL}
\end{equation}
Examining \eqref{gFriedmanHcombL}, we notice the special case where $\omega = -1$, resulting in a solution of $H=\text{const}$, akin to General Relativity. Since \eqref{gFriedmanHcombL} is a separable equation, we can express the \textbf{Type I} solution in the form of quadratures:
\begin{equation}
	\int \frac{\left( 2- N \right) +  \sum_{i=2}^{n}\alpha_i\prescript{(n)}{}k_{12}H^{2i-2} }{\frac{1}{2}\left( N - 1 \right)\left( N - 2 \right)H^2 - \sum_{i=2}^{n}\alpha_i \prescript{(n)}{}k_{21} H^{2i}  } dt = (1+\omega)t + C.
	\label{FlatLinGen}
\end{equation}

Unfortunately, it is impossible to write an explicit solution for \eqref{FlatLinGen} without fixing theory parameters such as $N$ - number of dimensions, $n$ - Lovelock polynomial order, and $\alpha_i$ - coupling constants. But it is clear that the extra term which exists in Lovelock gravity could play a Dark Energy component role.

\subsection{Open universe ($k=-1$) solutions}
For the open universe, it is advantageous to utilize the conformal flat form of the FRWL metric \cite{Landau1994} as given by:
\begin{equation}
    ds^2 = a(\eta)^2\left(-d\eta^2 + d\chi^2 + \sinh^2(\chi) d\Omega_{N-2}^2\right),
    \label{metricNf}
\end{equation}
where $dt = ad\eta$. The Riemann tensor components for the open universe metric \eqref{metricNf} are:
\begin{subequations}
    \begin{align}
        f_1 &= \frac{\dot{a}^2}{a^4} - \frac{1}{a^2},\\
        f_2 &= \frac{\ddot{a}}{a^3} - \frac{\dot{a}^2}{a^4},
    \end{align}
\end{subequations}
where dots indicate derivatives with respect to $\eta$. 

The Friedmann equations \eqref{RFriedman1}-\eqref{RFriedman2} for this scenario can be expressed as:
\begin{subequations}
    \begin{multline}
        -\frac{\left( N - 2 \right)}{2}\left( \left(N - 5\right)\frac{\dot{a}^2}{a^4} - \left(N - 3\right)\frac{1}{a^2} + 2\frac{\ddot{a}}{a^3} \right) +\\
        +\sum_{i=2}^n \alpha_i \prescript{(n)}{}k_{11}\left( \frac{\dot{a}^2}{a^4} - \frac{1}{a^2} \right)^i + \alpha_i \prescript{(n)}{}k_{12}\left( \frac{\dot{a}^2}{a^4} - \frac{1}{a^2} \right)^{i-1}\left( \frac{\ddot{a}}{a^3} - \frac{\dot{a}^2}{a^4} \right) = p,
        \label{OpenFriedman1}
    \end{multline}
    \begin{equation}
        -\frac{1}{2}\left( N - 1 \right)\left( N - 2 \right)\left( \frac{\dot{a}^2}{a^4} - \frac{1}{a^2} \right) + \sum_{i=2}^n \alpha_i \prescript{(n)}{}k_{21}\left( \frac{\dot{a}^2}{a^4} - \frac{1}{a^2} \right)^{i} = -\rho.
        \label{OpenFriedman2}
    \end{equation}
\end{subequations}
The field equations \eqref{OpenFriedman1}-\eqref{OpenFriedman2} reveal another well-known trivial vacuum solution in Lovelock gravity:
\begin{align}
    a &= Ce^{\eta},\\
    t &= C\left(e^{\eta}-1\right),
\end{align}
or equivalently as:
\begin{equation}
    \boxed{
    a = t,
}
\label{MilneSol}
\end{equation}
where we omit the constant related to the time definition $C$ in the $a = t+C$ solution. This is recognized as the \textbf{Type II} solution. This solution is also present in classic GR. From a physical point of view, it means that spacetime behavior in vacuum follows only negative Gaussian curvature.

\subsubsection{Pressure-free ($p=0$) solution}
Considering equation \eqref{OpenFriedman1}, we obtain:
\begin{multline} -\frac{\left( N - 2 \right)}{2}\left( \left(N - 5\right)\frac{\dot{a}^2}{a^4} - \left(N - 3\right)\frac{1}{a^2} + 2\frac{\ddot{a}}{a^3} \right) +\\
    +\sum_{i=2}^n \alpha_i \prescript{(n)}{}k_{11}\left( \frac{\dot{a}^2}{a^4} - \frac{1}{a^2} \right)^i + \alpha_i \prescript{(n)}{}k_{12}\left( \frac{\dot{a}^2}{a^4} - \frac{1}{a^2} \right)^{i-1}\left( \frac{\ddot{a}}{a^3} - \frac{\dot{a}^2}{a^4} \right) = 0.
    \label{PFopen1}
\end{multline}
To simplify \eqref{PFopen1}, let's introduce the transformation $z=\dot{a}$:
\begin{multline}
    \left[ -\left( N - 2 \right) + \sum_{i=2}^n \alpha_i \prescript{(n)}{}k_{12}\left( \frac{z^2}{a^4} - \frac{1}{a^2} \right)^{i - 1} \right]\frac{zz'}{a^3} + \\
    -\frac{\left( N - 2 \right)}{2}\left( \left(N - 5\right)\frac{z^2}{a^4} - \left(N - 3\right)\frac{1}{a^2} \right) + \sum_{i=2}^n  \alpha_i \prescript{(n)}{}k_{11}\left( \frac{z^2}{a^4} - \frac{1}{a^2} \right)^i - \alpha_i \prescript{(n)}{}k_{12}\frac{z^2}{a^4}\left( \frac{z^2}{a^4} - \frac{1}{a^2} \right)^{i-1} = 0.
    \label{PFopen2}
\end{multline}

Further manipulation using \eqref{eqK11}-\eqref{eqK22} simplifies \eqref{PFopen2} to:
\begin{equation}
    \sum_{i=1}^n \alpha_i \prescript{(n)}{}k_{12}\left( \frac{z^2}{a^4} - \frac{1}{a^2} \right)^{i-1}\left( z' + \frac{N - 4i - 1}{2i}\frac{z}{a} - \frac{N - 2i - 1}{2i}\frac{a}{z} \right) = 0,
    \label{PFopen3}
\end{equation}
where $\alpha_1 = 1$. While integrating \eqref{PFopen3} is non-trivial, it follows the form $Q(z,a)da - P(z,a)dz = 0$, which is well-studied in Lie point methods for solving differential equations (refer to \cite{Ibragimov1999}). Let's focus on a particular solution using the ansatz:
\begin{equation}
    z^2 = a^2\left( 1 + C_1 a^2 \right),
    \label{PFopen4}
\end{equation}
aiming to simplify the terms in \eqref{PFopen3}. Substituting \eqref{PFopen4} into \eqref{PFopen3} yields:
\begin{equation}
    \sum_{i=1}^n \alpha_i \prescript{(n)}{}k_{12} \frac{C_1^i}{i} = 0,
    \label{PFopen5}
\end{equation}
where we have already explored the case $C_1 = 0$ (Milne universe). Therefore, $C_1$ can be determined as a root of an algebraic polynomial equation:
\begin{equation}
    C_1 = \text{RootOf}\left( 2-N +\sum_{i=2}^n \alpha_i \prescript{(n)}{}k_{12}\frac{C^{i-1}_1}{i} = 0 \right).
    \label{PFopen6}
\end{equation}

It is evident from \eqref{PFopen6} that this solution only exists in Lovelock gravity ($n>1$). Consequently, the solution is:
\begin{align}
    \label{PFopen6.1}
    a &= \pm\frac{1}{\sqrt{C_1}\sinh\left( C_2 - \eta \right)},\\
    t &= \pm\frac{2}{\sqrt{C_1}}\arctanh e^{C_2 - \eta} + C_3.
    \label{PFopen6.2}
\end{align}
After simplification, we arrive at:
\begin{equation}
    a = \frac{1}{\sqrt{C_1}}\sinh\left( \sqrt{C_1}\left( t -  C_3\right) \right),
    \label{tmp1}
\end{equation}
Setting $C_3=0$ to satisfy $a(0)=0$, we obtain:
\begin{equation}
    \begin{aligned}
        a & = \frac{1}{\sqrt{C_1}}\sinh\left( \sqrt{C_1} t \right),\\
        C_1 &= \text{RootOf}\left( 2-N +\sum_{i=2}^n \alpha_i \prescript{(n)}{}k_{12}\frac{C^{i-1}_1}{i} = 0 \right).
    \end{aligned}
    \label{PFopen7}
\end{equation}
The solution \eqref{PFopen7} represents the Type I solution. It can be demonstrated that the solution $a = \frac{1}{C}\sinh\left( Ct \right)$ can exist in General Relativity only for the equation of state $p = -\rho$. From a physical point of view, \eqref{PFopen7} shows that Lovelock gravity provides an extra term which plays the role of a Dark Energy component.

\subsection{Closed universe ($k=1$) solutions}
For the closed universe, the conformal flat form of the FRWL metric is recommended as given by \cite{Landau1994}:
\begin{equation}
    ds^2 = a(\eta)^2\left(-d\eta^2 + d\chi^2 + \sin^2(\chi) d\Omega_{N-2}^2\right),
    \label{metricPf}
\end{equation}
where $dt = ad\eta$. The Riemann tensor components for the closed universe metric \eqref{metricPf} are:

\begin{subequations}
    \begin{align}
        f_1 &= \frac{\dot{a}^2}{a^4} + \frac{1}{a^2},\\
        f_2 &= \frac{\ddot{a}}{a^3} - \frac{\dot{a}^2}{a^4},
    \end{align}
\end{subequations}
where dots now signify derivatives with respect to $\eta$.

The Friedmann equations \eqref{RFriedman1}-\eqref{RFriedman2} in this scenario are:
\begin{subequations}
    \begin{multline}
        -\frac{\left( N - 2 \right)}{2}\left( \left(N - 5\right)\frac{\dot{a}^2}{a^4} + \left(N - 3\right)\frac{1}{a^2} + 2\frac{\ddot{a}}{a^3} \right) +\\
        +\sum_{i=2}^n \alpha_i \prescript{(n)}{}k_{11}\left( \frac{\dot{a}^2}{a^4} + \frac{1}{a^2} \right)^i + \alpha_i \prescript{(n)}{}k_{12}\left( \frac{\dot{a}^2}{a^4} + \frac{1}{a^2} \right)^{i-1}\left( \frac{\ddot{a}}{a^3} - \frac{\dot{a}^2}{a^4} \right) = p,
        \label{ClosedFriedman1}
\end{multline}
\begin{equation}
        -\frac{1}{2}\left( N - 1 \right)\left( N - 2 \right)\left( \frac{\dot{a}^2}{a^4} + \frac{1}{a^2} \right) + \sum_{i=2}^n \alpha_i \prescript{(n)}{}k_{21}\left( \frac{\dot{a}^2}{a^4} + \frac{1}{a^2} \right)^{i} = -\rho.
        \label{ClosedFriedman2}
\end{equation}
\end{subequations}

The methodology for obtaining solutions in a closed universe is analogous to that of the open universe.

\subsubsection{Pressure-free ($p=0$) solution}
Starting with equation \eqref{ClosedFriedman1}, we have:
\begin{multline}
    -\frac{\left( N - 2 \right)}{2}\left( \left(N - 5\right)\frac{\dot{a}^2}{a^4} + \left(N - 3\right)\frac{1}{a^2} + 2\frac{\ddot{a}}{a^3} \right) +\\
    +\sum_{i=2}^n \alpha_i \prescript{(n)}{}k_{11}\left( \frac{\dot{a}^2}{a^4} + \frac{1}{a^2} \right)^i + \alpha_i \prescript{(n)}{}k_{12}\left( \frac{\dot{a}^2}{a^4} + \frac{1}{a^2} \right)^{i-1}\left( \frac{\ddot{a}}{a^3} - \frac{\dot{a}^2}{a^4} \right) = 0.
    \label{PFclose1}
\end{multline}
To simplify \eqref{PFclose1}, we introduce the transformation $z=\dot{a}$:
\begin{multline}
    \left[ -\left( N - 2 \right) + \sum_{i=2}^n \alpha_i \prescript{(n)}{}k_{12}\left( \frac{z^2}{a^4} + \frac{1}{a^2} \right)^{i - 1} \right]\frac{zz'}{a^3} + \\
    -\frac{\left( N - 2 \right)}{2}\left( \left(N - 5\right)\frac{z^2}{a^4} + \left(N - 3\right)\frac{1}{a^2} \right) + \sum_{i=2}^n  \alpha_i \prescript{(n)}{}k_{11}\left( \frac{z^2}{a^4} + \frac{1}{a^2} \right)^i - \alpha_i \prescript{(n)}{}k_{12}\frac{z^2}{a^4}\left( \frac{z^2}{a^4} + \frac{1}{a^2} \right)^{i-1} = 0.
    \label{PFclose2}
\end{multline}
Utilizing \eqref{eqK11}-\eqref{eqK22}, we simplify \eqref{PFclose2} to:
\begin{equation}
    \sum_{i=1}^n \alpha_i \prescript{(n)}{}k_{12}\left( \frac{z^2}{a^4} + \frac{1}{a^2} \right)^{i-1}\left( z' + \frac{N - 4i - 1}{2i}\frac{z}{a} + \frac{N - 2i - 1}{2i}\frac{a}{z} \right) = 0,
    \label{PFclose3}
\end{equation}
where $\alpha_1 = 1$. Equation \eqref{PFopen3} is challenging to integrate, yet we can obtain a particular solution using an ansatz that simplifies the bracket within the sum:
\begin{equation}
    z^2 = a^2\left( C_1 a^2 - 1 \right),
    \label{PFcolse4}
\end{equation}
Upon replacing \eqref{PFcolse4} into \eqref{PFclose3}, we arrive at:
\begin{equation}
    \sum_{i=1}^n\alpha_i \prescript{(n)}{}k_{12}\frac{C_1^i}{i} = 0
    \label{PFclose5}
\end{equation}
which mirrors the expression in \eqref{PFopen5}. The solution in parametric form is:
\begin{align}
    a &= \pm\frac{1}{\sqrt{C_1}\sin\left( C_2 - \eta \right)},\\
    t &= \mp \frac{1}{\sqrt{C_1}}\ln \left( \frac{1 - \cos(C_2 - \eta)}{\sin\left( C_2 - \eta \right)} \right) + C_3,
\end{align}
and after some simplification is:
\begin{equation}
    a = \pm \cosh\left( \sqrt{C_1}\left( t - C_3 \right) \right),
\end{equation}
Consequently, the solution is represented as:
\begin{equation}
    \boxed{
    \begin{aligned}
        a & = \frac{1}{\sqrt{C_1}}\cosh\left( \sqrt{C_1}t \right),\\
        C_1 &= RootOf\left( 2-N +\sum_{i=2}^n \alpha_i \prescript{(n)}{}k_{12}\frac{C^{i-1}_1}{i} = 0 \right).
    \end{aligned}
}
\label{PFclose6}
\end{equation}
A notable observation is that there is no limit for $a(0)=0$ in this case. The solution \eqref{PFclose6} is considered the Type I solution. Similar to the open universe case, it can be demonstrated that the solution $a = \frac{1}{C}\cosh(Ct)$ can only be obtained in General Relativity for the equation of state $p=-\rho$. From a physical point of view, we see again that Lovelock gravity in general provides an extra term which plays a role of a Dark Energy component.

\section{Lovelock cosmology compared to Einstein cosmology}
In this section, we compare cosmology in Lovelock gravity with General Relativity cosmology without the lambda term in general. The LambdaCDM model is good for cosmology in practical point of view. But it has a Dark Energy problem, in other words, it needs an extra term leading to a negative equation of state ($p=-\rho$) which has no physical explanation. Thus, from modified gravitational theories, we expect to represent LambdaCDM model results but without involving Dark Energy.

Lovelock cosmology has extra freedom compared to classical General Relativity, namely $N$ - number of dimensions ($N=4$ in GR), $n$ - polynomial order ($n=1$ in GR), and $\alpha_i$ - coupling constants ($\alpha_i=0$ in GR). Without fixing these parameters, we can see that trivial GR vacuum solutions exist in Lovelock cosmology (namely \eqref{Mink}, \eqref{MilneSol}). The discovered feature of Lovelock cosmology is the existence of expanding vacuum solutions in general~\eqref{th1eq}. Which means that Lovelock cosmology provides Dark Energy behavior by design.

We should also notice that in Lovelock cosmology, in the presence of a perfect fluid matter distribution, it is possible to obtain solutions which correspond to General Relativity solutions: \eqref{FlatLinGen} for flat, \eqref{PFopen7} for open (pressure-free case), and \eqref{PFclose6} for closed (pressure-free case). To reproduce GR solution, one needs to set $N=4$, $n=1$, and $\alpha_i=0$ in the obtained results. Thus, Lovelock cosmology has extra freedom which does not exist in GR, and theory parameters could be used to reproduce Dark Energy behavior.

The main result of this paper is the Lovelock cosmology field equations in terms of Riemann tensor components \eqref{RFriedman1}-\eqref{RFriedman2}. Which represents the connection of spacetime curvature with matter. From these equations, one can easily resume Friedmann equations by setting $N=4$, $n=1$, and $\alpha_i=0$. In Lovelock gravity, cosmological equations show that spacetime curvature is connected to matter via algebraic equations like in GR but in a more complicated way. Unlike other modified theories of gravity (for instance, $f(R)$ gravity), where there is no extra derivation on Riemann tensor components in field equations, which follows the spirit of General Relativity.

\section{Conclusions}
In this paper, we have expanded upon the methodology initially introduced by Karmarkar to examine the embedding condition for Karmarkar class I spacetimes. Our extension encompasses $N$-dimensional spherically symmetric spacetimes and applies this approach to Lovelock gravity, establishing a robust framework for analyzing the field equations utilizing the Riemann tensor components within such contexts.

We introduced a notation based on the independent components of the Riemann tensor, enabling us to streamline and elucidate the analysis of the Lovelock field equations. By formulating the field equations in these terms, we deduced generalized Friedmann equations for Lovelock gravity that are relevant for both flat and open universes. Our solutions were classified into Type I and Type II, underscoring the presence of both trivial and non-trivial vacuum solutions within Lovelock gravity.

In the context of a flat universe, we uncovered additional vacuum solutions beyond those established in General Relativity. For an open universe, we derived a comprehensive solution for a pressure-free scenario unique to Lovelock gravity, absent in General Relativity. Similarly, in the case of a closed universe, we obtained a Type I solution exclusive to Lovelock gravity.

Our study offers a methodical and intricate examination of the Friedmann field equations within the realm of Lovelock gravity, introducing fresh perspectives on the solutions and their cosmological implications. The findings presented herein pave the way for further exploration into the cosmological ramifications of Lovelock gravity, potentially furnishing a more exhaustive comprehension of the universe's evolutionary trajectory.

The suggested approach could be leveraged to derive Lovelock field equations for a more general case of a spherically symmetric metric, as elaborated in forthcoming work.

The article was written within the framework of Additional Agreement No. 073-03-2024-
060/1 dated February 13, 2024 to the Agreement on the provision of subsidies from the federal
budget for financial support for the implementation of the state task for the provision of public
services (performance of work) No. 073-03-2024-060 dated January 18, 2024, concluded between
the Federal State Budgetary Educational Institution of Higher Education “UlSPU I.N. Ulyanov”
and the Ministry of Education of the Russian Federation.

\bibliographystyle{plain}
\bibliography{bibl}

\appendix

\section{Riemann tensor in $N$-dimensional spherical symmetric spacetime}
\label{app1}
In this appendix our goal is to proof \eqref{notation5_f}-\eqref{notation5_l} which we made by guess
\begin{subequations}
    \begin{align}
        \label{app1_eq1}
        f_1 =& R^{1\alpha}_{\ \ 1\alpha} = R^{\alpha\beta}_{\ \ \alpha\beta} = \frac{\dot{a}^2}{a^2} + \frac{k}{a^2},\\
        f_2 =& R^{10}_{\ \ 10} = R^{\alpha 0}_{\ \ \alpha 0} = \frac{\ddot{a}}{a}.
        \label{app1_eq2}
    \end{align}
\end{subequations}

We will use the definition of the Riemann tensor \cite{Landau1994}
\begin{equation}
    R^{ab}_{\ \ cd} = g^{be}R^a_{\ ecd} = g^{be}\left[ \Gamma^a_{ed,c} - \Gamma^a_{ec,d} + \Gamma^a_{fc}\Gamma^f_{ed} - \Gamma^a_{gd}\Gamma^g_{ec} \right],
    \label{RiemannDef}
\end{equation}
where $\Gamma$ is the Christoffel symbol of second type
\begin{equation}
    \Gamma^a_{bc} = \frac{1}{2}g^{ad}\left( g_{db,c} + g_{dc,b} - g_{bc,d} \right).
    \label{ChrisoffelDef}
\end{equation}

First we will proof
\begin{equation}
    f_1 = R^{1\alpha}_{\ \ 1\alpha},
    \label{app1_eq3}
\end{equation}
from \eqref{RiemannDef} we have
\begin{multline}
    R^{1\alpha}_{\ \ 1\alpha} = g^{\alpha\alpha}\left[ \Gamma^{1}_{\alpha\alpha,1} - \Gamma^1_{\alpha 1,\alpha} + \Gamma^1_{a1}\Gamma^a_{\alpha\alpha} - \Gamma^1_{b\alpha}\Gamma^b_{\alpha 1} \right] = \\
    g^{\alpha\alpha}\left[ -\frac{1}{2}\left( g^{11}g_{\alpha\alpha,1} \right)_{,1} + \frac{1}{4}g^{11}g^{aa}\left( g_{1a,1} + g_{11,a} - g_{a1,1} \right)\left( g_{a\alpha,\alpha} + g_{a\alpha,\alpha} - g_{\alpha\alpha,a} \right) \right. - \\
    \left. -\frac{1}{4}g^{11}g^{bb}\left( g_{1b,\alpha} + g_{1\alpha,b} - g_{b\alpha,1} \right)\left( g_{b\alpha,1} + g_{b1,\alpha} - g_{\alpha 1,b} \right)  \right] =\\
    -\frac{1}{r}g^{11}_{,1} - \frac{1}{r^2}g^{11} - \frac{\dot{a}}{2a}g^{11}g^{00}g_{11,0} -\frac{1}{2r}\left( g^{11} \right)^2g_{11,1} + \frac{1}{r^2}g^{11} = \frac{\dot{a}^2}{a^2} + \frac{k}{a^2}
    \label{app1_eq4}
\end{multline}
where we used $g_{1\alpha} = 0$, $g_{11,\alpha} = 0$, $g^{\alpha\alpha}g_{\alpha\alpha,1} = \frac{2}{r}$, $g^{\alpha\alpha}g_{\alpha\alpha,11} = \frac{2}{r^2}$, $g^{\alpha\alpha}g_{\alpha\alpha,0} = \frac{2\dot{a}}{a}$. Thus \eqref{app1_eq3} is proofed.

Next we will proof
\begin{equation}
    f_1 = R^{\alpha\beta}_{\ \ \alpha\beta},
    \label{app1_eq5}
\end{equation}
where $\beta > \alpha$. From \eqref{RiemannDef} we have
\begin{multline}
    R^{\alpha\beta}_{\ \ \alpha\beta} = g^{\beta\beta}\left( \Gamma^\alpha_{\beta\beta,\alpha} - \Gamma^{\alpha}_{\beta\alpha,\beta} + \Gamma^\alpha_{a\alpha}\Gamma^a_{\beta\beta} - \Gamma^\alpha_{b\beta}\Gamma^b_{\beta\alpha} \right) = \\
    - \frac{1}{2}g^{\alpha\alpha}g^{\beta\beta}g_{\beta\beta,\alpha\alpha} + \frac{1}{4}g^{\beta\beta}g^{\alpha\alpha}g^{aa}\left( g_{\alpha a,\alpha} + g_{\alpha\alpha, a} - g_{a\alpha,\alpha} \right)\left( g_{a\beta,\beta} + g_{a\beta,\beta} - g_{\beta\beta,a} \right) - \\
    -\frac{1}{4}g^{\beta\beta}g^{\alpha\alpha}g^{bb}\left( g_{\alpha b,\beta} + g_{\alpha\beta,b} - g_{b\beta,\alpha} \right)\left( g_{b\beta,\alpha} + g_{b\alpha,\beta} - g_{\beta\alpha,b} \right) =\\
    - \frac{1}{2}g^{\alpha\alpha}g^{\beta\beta}g_{\beta\beta,\alpha\alpha} - \frac{\dot{a}^2}{a^2}g^{00} - \frac{1}{r^2}g^{11} -\frac{1}{4}g^{\beta\beta}g^{\alpha\alpha}\sum_{\gamma = 2}^{\alpha - 1}g^{\gamma\gamma}g_{\alpha\alpha,\gamma}g_{\beta\beta,\gamma} + \frac{1}{4}g^{\alpha\alpha}g^{\beta\beta}g^{\beta\beta}g_{\beta\beta,\alpha}g_{\beta\beta,\alpha} = \\
    -g^{\alpha\alpha}\left( \cot^2\phi_\alpha - 1 \right) - \frac{\dot{a}^2}{a^2}g^{00} - \frac{1}{r^2}g^{11} - \sum_{\gamma = 2}^{\alpha - 1}g^{\gamma\gamma}\frac{\cos^2\phi_{\gamma-1}}{\sin^2\phi_{\gamma-1}} + g^{\alpha\alpha}\cot^2\phi_\alpha = g^{22} - \frac{\dot{a}^2}{a^2}g^{00} - \frac{1}{r^2}g^{11} = \frac{\dot{a}^2}{a^2} + \frac{k}{a^2},
    \label{app1_eq6}
\end{multline}
where we used $g_{\alpha\beta} = 0$, $g^{\alpha\alpha}_{,\alpha} = 0$, $g_{\alpha\alpha,\beta} = 0$, $g^{\beta\beta}g_{\beta\beta,\alpha} = 2\cot\phi_\alpha$, $g^{\beta\beta}g_{\beta\beta,\alpha\alpha} = 2\left( \cot^2\phi_\alpha - 1 \right)$ and
\begin{equation}
    g^{\alpha\alpha} - \sum_{\gamma = 2}^{\alpha - 1} g^{\gamma\gamma}\frac{\cos^2 \phi_{\gamma-1}}{\sin^2 \phi_{\gamma-1}} = g^{22}.
    \label{app1_eq7}
\end{equation}
Thus we proofed \eqref{app1_eq5}.
The above statement \eqref{app1_eq7} easy to proof using method of mathematical induction
\begin{enumerate}
    \item It is true for $\alpha = 3$
        \begin{equation}
            g^{33} - g^{22}\frac{\cos^2 \phi_1}{\sin^2 \phi_1} = g^{33} - g^{33}\cos^2 \phi_1 = g^{33}\sin^2 \phi_1 = g^{22}.
            \label{app1_eq8}
        \end{equation}
    \item If it is true for $\alpha = n > 3$
        \begin{equation}
            g^{nn} - \sum_{\gamma = 2}^{n-1} g^{\gamma\gamma}\frac{\cos^2 \phi_{\gamma-1}}{\sin^2 \phi_{\gamma-1}} = g^{22},
            \label{app1_eq9}
        \end{equation}
        then for $\alpha = n + 1$ it is
        \begin{multline}
            g^{n+1n+1} - g^{nn}\frac{\cos^2\phi_{n-1}}{\sin^2\phi_{n-1}} -  \sum_{\gamma = 2}^{n-1} g^{\gamma\gamma}\frac{\cos^2 \phi_{\gamma-1}}{\sin^2 \phi_{\gamma-1}} = \\
            g^{n+1n+1} - g^{n+1n+1}\cos^2\phi_{n-1} -  \sum_{\gamma = 2}^{n-1} g^{\gamma\gamma}\frac{\cos^2 \phi_{\gamma-1}}{\sin^2 \phi_{\gamma-1}} = g^{nn} - \sum_{\gamma = 2}^{n-1} g^{\gamma\gamma}\frac{\cos^2 \phi_{\gamma-1}}{\sin^2 \phi_{\gamma-1}}  = g^{22},
            \label{app1_eq10}
        \end{multline}
        which is true. As a result we proofed \eqref{app1_eq7}.
\end{enumerate}

The final step is to proof
\begin{equation}
    f_2 = R^{10}_{\ \ 10} = R^{\alpha 0}_{\ \ \alpha 0}.
    \label{app1_eq11}
\end{equation}
Using definition of the Riemann tensor
\begin{subequations}
    \begin{multline}
        R^{\alpha 0}_{\ \ \alpha 0} = g^{00}\left[ \Gamma^\alpha_{00,\alpha} - \Gamma^\alpha_{0\alpha,0} + \Gamma^{\alpha}_{a\alpha}\Gamma^a_{00} - \Gamma^\alpha_{b0}\Gamma^b_{0\alpha} \right] = \\
        - \frac{1}{2}g^{00}\left( g^{\alpha\alpha}g_{\alpha\alpha,0} \right)_{,0} + \frac{1}{4}g^{00}g^{\alpha\alpha}g^{aa}g_{\alpha\alpha,a}\left( g_{a0,0} + g_{a0,a} - g_{00,a} \right) -\\
        -\frac{1}{4}g^{00}g^{\alpha\alpha}g^{bb} g_{\alpha b,0} g_{b\alpha,0} = -g^{00}\left( \frac{\ddot{a}}{a} - \frac{\dot{a}^2}{a^2} \right) - g^{00}\frac{\dot{a}^2}{a^2} = \frac{\ddot{a}}{a}
        \label{app1_eq12}
    \end{multline}
    \begin{multline}
        R^{10}_{\ \ 10} = g^{00}\left[ \Gamma^1_{00,1} - \Gamma^1_{01,0} + \Gamma^{1}_{a1}\Gamma^a_{00} - \Gamma^1_{b0}\Gamma^b_{01} \right] = \\
        -\frac{1}{2}g^{00}\left( g^{11}g_{11,0} \right)_{,0} - \frac{1}{4}g^{00}g^{11}g^{11}g_{11,0}g_{11,0} = \left( \frac{\ddot{a}}{a} - \frac{\dot{a}^2}{a^2} \right) +\frac{\dot{a}^2}{a^2} = \frac{\ddot{a}}{a}
    \end{multline}
\end{subequations}
where we used $g_{00,\alpha} = 0$, $g^{\alpha\alpha}g_{\alpha\alpha,0} = \frac{2\dot{a}}{a}$, $g_{00,a} = 0$, $g^{11}g_{11,0} = \frac{2\dot{a}}{a}$.

\section{Friedmann equations for Lovelock gravity}
\label{app2}
To simplify practical usage of obtained generalized Friedmann equations \eqref{gFriedman1}-\eqref{gFriedman2} we write down it's particular cases.

The First order Lovelock polynomial Friedmann equations are
\begin{subequations}
    \begin{equation}
        -\left( N - 2 \right)\left( \frac{\ddot{a}}{a} + \frac{N-3}{2}\left( \frac{\dot{a}^2}{a^2} + \frac{k}{a^2} \right) \right) = p,
        \label{1Friedmann11}
    \end{equation}
    \begin{equation}
        -\frac{\left( N - 1 \right)\left( N - 2 \right)}{2}\left( \frac{\dot{a}^2}{a^2} + \frac{k}{a^2} \right) = -\rho,
        \label{1Friedmann00}
    \end{equation}
\end{subequations}
which are $N$-dimensional Friedmann equations. Equations for particular $N$ are given in table \ref{tab2}.

\begin{table}[h!]
    \centering
    \renewcommand{\arraystretch}{1.3}
    \begin{tabular}{ l  c }
        \hline
        $N$ & Friedmann equations ($n=1$) \\ [0.5ex]
        \hline\hline
        $4$ & $\begin{array} {lcl}  -2\frac{\ddot{a}}{a} -\frac{\dot{a}^2}{a^2} - \frac{k}{a^2}  =  p \\ -3\frac{\dot{a}^2}{a^2} - 3\frac{k}{a^2}  =  -\rho \end{array}$ \\
        \hline
        $5$ & $\begin{array} {lcl} -3\frac{\ddot{a}}{a} -3\frac{\dot{a}^2}{a^2} - 3\frac{k}{a^2}  =  p \\ -6\frac{\dot{a}^2}{a^2} - 6\frac{k}{a^2}  =  -\rho \end{array}$ \\
        \hline
        $6$ & $\begin{array} {lcl}  -4\frac{\ddot{a}}{a} -6\frac{\dot{a}^2}{a^2} - 6\frac{k}{a^2}  =  p \\ -10\frac{\dot{a}^2}{a^2} - 10\frac{k}{a^2}  =  -\rho \end{array}$ \\
        \hline
        $7$ & $\begin{array} {lcl} -5\frac{\ddot{a}}{a} -10\frac{\dot{a}^2}{a^2} - 10\frac{k}{a^2}  =  p \\ -15\frac{\dot{a}^2}{a^2} - 15\frac{k}{a^2}  =  -\rho \end{array}$ \\
        \hline
        $8$ & $\begin{array} {lcl} -6\frac{\ddot{a}}{a} -15\frac{\dot{a}^2}{a^2} - 15\frac{k}{a^2}  =  p \\ -21\frac{\dot{a}^2}{a^2} - 21\frac{k}{a^2}  =  -\rho \end{array}$ \\
        \hline
        $9$ & $\begin{array} {lcl} -7\frac{\ddot{a}}{a} -21\frac{\dot{a}^2}{a^2} - 21\frac{k}{a^2}  =  p \\ -28\frac{\dot{a}^2}{a^2} - 28\frac{k}{a^2}  =  -\rho \end{array}$ \\
        \hline
        $10$ & $\begin{array} {lcl} -8\frac{\ddot{a}}{a} -28\frac{\dot{a}^2}{a^2} - 28\frac{k}{a^2}  =  p \\ -36\frac{\dot{a}^2}{a^2} - 36\frac{k}{a^2}  =  -\rho \end{array}$ \\
        \hline
        \hline
    \end{tabular}
    \caption{Friedmann equations for first order Lovelock gravity (General relativity) $n=1$}
    \label{tab2}
\end{table}

The Second order Lovelock polynomial Friedmann equations are
\begin{subequations}
    \begin{multline}
        -\left( N - 2 \right)\left( \frac{\ddot{a}}{a} + \frac{N-3}{2}\left( \frac{\dot{a}^2}{a^2} + \frac{k}{a^2} \right) \right) -\\- \alpha \frac{\left( N - 2 \right)\left( N - 3 \right)\left( N - 4 \right)}{2}\left[ 4\frac{\ddot{a}}{a} + \left( N - 5 \right)\left( \frac{\dot{a}^2}{a^2} + \frac{k}{a^2} \right)\right]\left( \frac{\dot{a}^2}{a^2} + \frac{k}{a^2} \right) = p,
        \label{2Friedmann11}
    \end{multline}
    \begin{multline}
        -\frac{\left( N - 1 \right)\left( N - 2 \right)}{2}\left( \frac{\dot{a}^2}{a^2} + \frac{k}{a^2} \right) - \alpha \frac{\left( N - 1 \right)\left( N - 2 \right)\left( N - 3 \right)\left( N - 4 \right)}{2}\left( \frac{\dot{a}^2}{a^2} + \frac{k}{a^2} \right)  = -\rho.
        \label{2Friedmann00}
    \end{multline}
\end{subequations}
Equations for particular $N$ are given in table \ref{tab3}.

\begin{table}[h!]
    \centering
    \renewcommand{\arraystretch}{1.3}
    \begin{tabular}{ l  c }
        \hline
        $N$ & Friedmann equations ($n=2$) \\ [0.5ex]
        \hline\hline
        $5$ & $\begin{array} {lcl} -3\frac{\ddot{a}}{a} -3\frac{\dot{a}^2}{a^2} - 3\frac{k}{a^2} -12\alpha \frac{\ddot{a}\left( \dot{a}^2 + k \right)}{a^3}  =  p \\ -6\frac{\dot{a}^2}{a^2} - 6\frac{k}{a^2} - 12\alpha \frac{\left(\dot{a}^2 + k\right)^2}{a^4}  =  -\rho \end{array}$ \\
        \hline
        $6$ & $\begin{array} {lcl} -4\frac{\ddot{a}}{a} -6\frac{\dot{a}^2}{a^2} - 6\frac{k}{a^2} -12 \alpha \frac{\left( 4\ddot{a}a + \dot{a}^2 + k \right)\left( \dot{a}^2 + k \right)}{a^4}  =  p \\ -10\frac{\dot{a}^2}{a^2} - 10\frac{k}{a^2} - 60\alpha \frac{\left( \dot{a}^2 + k \right)^2}{a^4}  =  -\rho \end{array}$ \\
        \hline
        $7$ & $\begin{array} {lcl} -5\frac{\ddot{a}}{a} -10\frac{\dot{a}^2}{a^2} - 10\frac{k}{a^2} -15 \alpha \frac{\left( 8\ddot{a}a + 4\dot{a}^2 + 4k \right)\left( \dot{a}^2 + k \right)}{a^4}   =  p \\ -15\frac{\dot{a}^2}{a^2} - 15\frac{k}{a^2} - 180\alpha \frac{\left( \dot{a}^2 + k \right)^2}{a^4}   =  -\rho \end{array}$ \\
        \hline
        $8$ & $\begin{array} {lcl} -6\frac{\ddot{a}}{a} -15\frac{\dot{a}^2}{a^2} - 15\frac{k}{a^2} -180 \alpha \frac{\left( \frac{4}{3}\ddot{a}a + \dot{a}^2 + k \right)\left( \dot{a}^2 + k \right)}{a^4}    =  p \\ -21\frac{\dot{a}^2}{a^2} - 21\frac{k}{a^2} - 420\alpha \frac{\left( \dot{a}^2 + k \right)^2}{a^4}  =  -\rho \end{array}$ \\
        \hline
        $9$ & $\begin{array} {lcl} -7\frac{\ddot{a}}{a} -21\frac{\dot{a}^2}{a^2} - 21\frac{k}{a^2}  -28 \alpha \frac{\left( 15\ddot{a}a + 15\dot{a}^2 + 15k \right)\left( \dot{a}^2 + k \right)}{a^4}   =  p \\ -28\frac{\dot{a}^2}{a^2} - 28\frac{k}{a^2} - 840\alpha \frac{\left( \dot{a}^2 + k \right)^2}{a^4}  =  -\rho \end{array}$ \\
        \hline
        $10$ & $\begin{array} {lcl} -8\frac{\ddot{a}}{a} -28\frac{\dot{a}^2}{a^2} - 28\frac{k}{a^2} -840 \alpha \frac{\left( \frac{4}{5}\ddot{a}a + \dot{a}^2 + k \right)\left( \dot{a}^2 + k \right)}{a^4}   =  p \\ -36\frac{\dot{a}^2}{a^2} - 36\frac{k}{a^2}  - 1512\alpha \frac{\left( \dot{a}^2 + k \right)^2}{a^4}   =  -\rho \end{array}$ \\
        \hline
        \hline
    \end{tabular}
    \caption{Friedmann equations for second order Lovelock gravity (EGB gravity) $n=2$}
    \label{tab3}
\end{table}

\begin{table}[h!]
    \centering
    \renewcommand{\arraystretch}{1.3}
    \begin{tabular}{ l  c }
        \hline
        $N$ & Friedmann equations ($n=3$) \\ [0.5ex]
        \hline\hline
        $7$ & $\begin{array} {lcl} -5\frac{\ddot{a}}{a} -10\frac{\dot{a}^2}{a^2} - 10\frac{k}{a^2} -15 \alpha \frac{\left( 8\ddot{a}a + 4\dot{a}^2 + 4k \right)\left( \dot{a}^2 + k \right)}{a^4} - 360\beta \frac{\ddot{a}\left( \dot{a}^2 + k \right)^2}{a^5}   =  p \\ -15\frac{\dot{a}^2}{a^2} - 15\frac{k}{a^2} - 180\alpha \frac{\left( \dot{a}^2 + k \right)^2}{a^4} - 360\beta \frac{\left( \dot{a} + k \right)^3}{a^6}    =  -\rho \end{array}$ \\
        \hline
        $8$ & $\begin{array} {lcl} -6\frac{\ddot{a}}{a} -15\frac{\dot{a}^2}{a^2} - 15\frac{k}{a^2} -180 \alpha \frac{\left( \frac{4}{3}\ddot{a}a + \dot{a}^2 + k \right)\left( \dot{a}^2 + k \right)}{a^4} -360\beta \frac{\left( 6\ddot{a}a + \dot{a}^2 + k \right)\left( \dot{a}^2 + k \right)^2}{a^6}   =  p \\ -21\frac{\dot{a}^2}{a^2} - 21\frac{k}{a^2} - 420\alpha \frac{\left( \dot{a}^2 + k \right)^2}{a^4} -2520\beta \frac{\left( \dot{a}^2 + k \right)^3}{a^6}  =  -\rho \end{array}$ \\
        \hline
        $9$ & $\begin{array} {lcl} -7\frac{\ddot{a}}{a} -21\frac{\dot{a}^2}{a^2} - 21\frac{k}{a^2}  -28 \alpha \frac{\left( 15\ddot{a}a + 15\dot{a}^2 + 15k \right)\left( \dot{a}^2 + k \right)}{a^4}  -2520\beta \frac{\left( 3\ddot{a}a + \dot{a}^2 + k \right)\left( \dot{a}^2 + k \right)^2}{a^6}    =  p \\ -28\frac{\dot{a}^2}{a^2} - 28\frac{k}{a^2} - 840\alpha \frac{\left( \dot{a}^2 + k \right)^2}{a^4}  -10080\beta \frac{\left( \dot{a}^2 + k \right)^3}{a^6}  =  -\rho \end{array}$ \\
        \hline
        $10$ & $\begin{array} {lcl} -8\frac{\ddot{a}}{a} -28\frac{\dot{a}^2}{a^2} - 28\frac{k}{a^2} -840 \alpha \frac{\left( \frac{4}{5}\ddot{a}a + \dot{a}^2 + k \right)\left( \dot{a}^2 + k \right)}{a^4}  -10080\beta \frac{\left( 2\ddot{a}a + \dot{a}^2 + k \right)\left( \dot{a}^2 + k \right)^2}{a^6}   =  p \\ -36\frac{\dot{a}^2}{a^2} - 36\frac{k}{a^2}  - 1512\alpha \frac{\left( \dot{a}^2 + k \right)^2}{a^4}  -30240\beta \frac{\left( \dot{a}^2 + k \right)^3}{a^6}   =  -\rho \end{array}$ \\
        \hline
        \hline
    \end{tabular}
    \caption{Friedmann equations for third order Lovelock gravity $n=3$}
    \label{tab4}
\end{table}

\begin{table}[h!]
    \centering
    \renewcommand{\arraystretch}{1.3}
    \begin{tabular}{ l  c }
        \hline
        $N$ & Friedmann equations ($n=4$) \\ [0.5ex]
        \hline\hline
        $9$ & $\begin{array} {lcl} -7\frac{\ddot{a}}{a} -21\frac{\dot{a}^2}{a^2} - 21\frac{k}{a^2}  -28 \alpha \frac{\left( 15\ddot{a}a + 15\dot{a}^2 + 15k \right)\left( \dot{a}^2 + k \right)}{a^4}  -2520\beta \frac{\left( 3\ddot{a}a + \dot{a}^2 + k \right)\left( \dot{a}^2 + k \right)^2}{a^6}  -\\-20160\gamma \frac{\ddot{a}\left( \dot{a}^2 + k \right)^3}{a^7}  =  p \\ -28\frac{\dot{a}^2}{a^2} - 28\frac{k}{a^2} - 840\alpha \frac{\left( \dot{a}^2 + k \right)^2}{a^4}  -10080\beta \frac{\left( \dot{a}^2 + k \right)^3}{a^6} - 20160\gamma \frac{\left( \dot{a}^2 + k \right)^4}{a^8}  =  -\rho \end{array}$ \\
        \hline
        $10$ & $\begin{array} {lcl} -8\frac{\ddot{a}}{a} -28\frac{\dot{a}^2}{a^2} - 28\frac{k}{a^2} -840 \alpha \frac{\left( \frac{4}{5}\ddot{a}a + \dot{a}^2 + k \right)\left( \dot{a}^2 + k \right)}{a^4}  -10080\beta \frac{\left( 2\ddot{a}a + \dot{a}^2 + k \right)\left( \dot{a}^2 + k \right)^2}{a^6} -\\- 20160\gamma\frac{\left( 8\ddot{a}a + \dot{a}^2 + k \right)\left( \dot{a}^2 + k \right)^3}{a^8}  =  p \\ -36\frac{\dot{a}^2}{a^2} - 36\frac{k}{a^2}  - 1512\alpha \frac{\left( \dot{a}^2 + k \right)^2}{a^4}  -30240\beta \frac{\left( \dot{a}^2 + k \right)^3}{a^6} - 181440\frac{\left(\dot{a}^2 + k\right)^4}{a^8}   =  -\rho \end{array}$ \\
        \hline
        \hline
    \end{tabular}
    \caption{Friedmann equations for fourth order Lovelock gravity $n=4$}
    \label{tab5}
\end{table}

\end{document}